\documentclass[aps,pra,superscriptaddress,floatfix,reprint,showkeys,showpacs]{revtex4-1}
\usepackage{graphicx}
\usepackage{ifpdf}

\ifpdf
	\usepackage{pdfpages}
	\usepackage{epstopdf}
\fi	

\usepackage{amsmath,amssymb,amstext}
\usepackage{color}
\usepackage{natbib}

\def\dd{\,\mathrm{d}}

\begin{document}
\title{Collective Excitation Interferometry with a Toroidal Bose-Einstein Condensate}
\author{G. Edward Marti}
\email{emarti@berkeley.edu}
\affiliation{Department of Physics, University of California, Berkeley, California 94720, USA}
\author{Ryan Olf}
\email{ryanolf@berkeley.edu}
\affiliation{Department of Physics, University of California, Berkeley, California 94720, USA}
\author{Dan M. Stamper-Kurn}
\affiliation{Department of Physics, University of California, Berkeley, California 94720, USA}
\affiliation{Materials Sciences Division, Lawrence Berkeley National Laboratory, Berkeley, California 94720, USA}

\date{\today}
\begin{abstract}
The precision of compact inertial sensing schemes using trapped- and guided-atom interferometers has been limited by uncontrolled phase errors caused by trapping potentials and interactions. Here, we propose an acoustic interferometer that uses sound waves in a toroidal Bose-Einstein condensate to measure rotation, and we demonstrate experimentally several key aspects of this type of interferometer.  We use spatially patterned light beams to excite counter-propagating sound waves within the condensate and use \emph{in situ} absorption imaging to characterize their evolution.  We present an analysis technique by which we extract separately the oscillation frequencies of the standing-wave acoustic modes, the frequency splitting caused by static imperfections in the trapping potential, and the characteristic precession of the standing-wave pattern due to rotation.  Supported by analytic and numerical calculations, we interpret the noise in our measurements, which is dominated by atom shot noise, in terms of rotation noise.  While the noise of our acoustic interferometric sensor, at the level of $\sim \mbox{rad}\,  \mbox{s}^{-1}/\sqrt{\mbox{Hz}}$, is high owing to rapid acoustic damping and the small radius of the trap, the proof-of-concept device does operate at $10^4 - 10^6$ times higher density and in a volume $10^9$ times smaller than free-falling atom interferometers.
\end{abstract}


\maketitle

Conventional atom interferometers measure acceleration~\cite{McGuirk2002}
and rotation~\cite{Gustavson1997,Gustavson2000, Durfee2006} by interfering dilute atomic wavepackets that traverse distinct paths in free fall~\cite{Cronin2009}.  The impressive sensitivity of these devices scales with the area enclosed by the arms of the interferometer, favoring larger interferometers that average measurements on centimeter length scales~\cite{Muller2008b}. Extending the capability of atom interferometers to probe shorter length scales could address fundamental questions, such as how gravity operates at short range; tackle practical problems, such as non-invasive material characterization; and aid in the development of miniaturized atomic sensors~\cite{Farkas2010}. Trapped- or guided-atom interferometers may allow sensitive, localized inertial measurements by allowing interferometer arms to enclose the same area multiple times, gaining precision while remaining compact. However, for an atom interferometer to reach high signal-to-noise and probe short length scales, it is critical to develop an interferometric scheme compatible with high densities and realistic trapping potentials.

Trapped quantum degenerate gases offer a bright source for atom interferometry in small volumes, reaching number densities of  $10^{14}\;\mathrm{cm}^{-3}$ that are at least four orders of magnitude higher than those utilized in free-falling-atom devices~\cite{Muller2008b}. However, the price of high density is uncontrolled phase shifts and damped atomic motion ~\cite{Wang2005, Garcia2006,Shin2004, Jo2007}. Also, in spite of high densities, the number flux of ultracold atoms through a trapped-atom interferometer is typically low, making it highly desirable that the readout noise of such interferometers reach, or even surpass~\cite{Esteve2008, Appel2009, Leroux2010, Schleier-Smith2010}, the atom-shot-noise limit.

Here, we propose and demonstrate a new type of interferometer that circumvents many challenges of high-density atom interferometry: interfering collective excitations of a dense, trapped sample to measure force or rotation. In this proof-of-principle work, we interfere phonons, our chosen collective excitation, in a toroidal Bose-Einstein condensate (BEC) and extract a signal that is sensitive to rotations. Our scheme is similar to those of hemispherical resonator~\cite{Rozelle2009} and superfluid-helium gyroscopes~\cite{Packard1992}.  In analogy to an optical gyroscope, phonons play the role of light, traveling through the vacuum mode of the BEC.  The effects of trap inhomogeneity and of atomic interactions are ameliorated in two ways. First, atomic interactions themselves are used to suppress the effect of trap inhomogeneity on sound propagation. We demonstrate this fact experimentally by showing that significant trap inhomogeneity leads only to weak coupling between counter propagating sound modes.  Second, we develop an analysis technique that allows us to isolate a rotation-sensitive signal from the dynamical evolution of sound waves in the toroidal BEC in a manner that is largely independent of the effects of interactions and of trap inhomogeneity.  This analysis technique is applied to experimental data to quantify the noise in a rotation-rate measurement, and to data generated by numerical simulations to quantify the rotation sensitivity.

We demonstrate several key advantages of collective-excitation interferometry by constructing a high-density ($1\times 10^{14}\;\mathrm{cm}^{-3}$) but small ($16\;\mu\mathrm{m}$ radius) sample. As collective modes of an interacting Bose-Einstein condensate, sound waves can, in principle, propagate over long distances. Interactions in a superfluid can enhance the lifetime of the sound mode even in the presence of disorder~\cite{Dries2010} and suppress systematic biases that arise from weak disorder in the potential.  Time reversal symmetry guarantees that linear forces, atom number variations, interaction energy shifts, and static trap inhomogeneities cannot distinguish counter-propagating acoustic modes; these effects primarily introduce common-mode phase shifts that do not deteriorate the signal. Further, the irrotational nature of the superfluid provides a non-rotating frame for the propagating sound waves, against which the slow rotation of an observer can be measured absolutely.  Our compact device presently achieves a rotation sensitivity of only ${\sim}\;\textrm{rad s}^{-1}/\sqrt{\mathrm{Hz}}$, far inferior to available sensors.  Extending our scheme to circular waveguides of millimeter dimensions~\cite{Gupta2005} and reducing damping to gain longer propagation times would be necessary to improve sensitivity.

Low-order collective modes of BEC's have been used to measure Casimir-Polder forces~\cite{Harber2005} and quantized circulation~\cite{Leanhardt2002a,Cozzini2003,Bretin2003}. We extend this work by using higher-order standing-wave acoustic modes to increase sensitivity, overcome technical noise limitations, and reach atom-shot-noise-limited detection. Since we lack the sensitivity to measure rotation directly, the goal of this paper is to validate our proposal by matching the noise of the rotation signal to the expected atom shot noise and by identifying systematic biases. The most critical bias we investigate results from azimuthal perturbations in the trap potential that add frequency shifts to the standing-wave eigenmodes, which could appear as rotation signal if not properly accounted for. We characterize and correct the rotational signal for these effects.

We begin in Sec.\ \ref{sec:setup} by describing our experimental system for producing toroidal-shaped Bose-Einstein condensed gases of $^{87}$Rb, our optical method for exciting standing-wave acoustic modes of various angular orders and initial angular positions, and our measurement of the evolution of these modes through \emph{in-situ} absorption imaging.  We characterise the frequency and spatial pattern of several acoustic collective modes.  We also observe these modes to be damped more rapidly than expected based on Landau damping.  In Sec.\ \ref{sec:epoc}, we show how the separate effects of static trap inhomogeneities and of rotation can be isolated in our data analysis.  The resulting rotation-rate measurements, based on the precession of standing-wave modes of different orders, are presented.  Because we lack sensitivity to measure rotation rates that we can reasonably apply to our experiment, we quantify the variance in these measurements as the noise in a rotation-rate measurement.  For rotation sensing using higher-order acoustic modes, we find this measurement noise to be consistent with that expected for atom-shot-noise-limited measurements.   Finally, in Section~\ref{sec:sims} we examine the limits of our scheme in the zero-temperature, mean-field limit by numerically simulating a representation of our system in the presence of rotation. In concert, this experimental and numerical work highlights many of the advantages of collective-excitation interferometry and can serve to guide future work in this area.

\section{Creation and Evolution of Sound Waves}\label{sec:setup}

\begin{figure}[tb!]
\begin{center}
\includegraphics[width=0.95 \columnwidth]{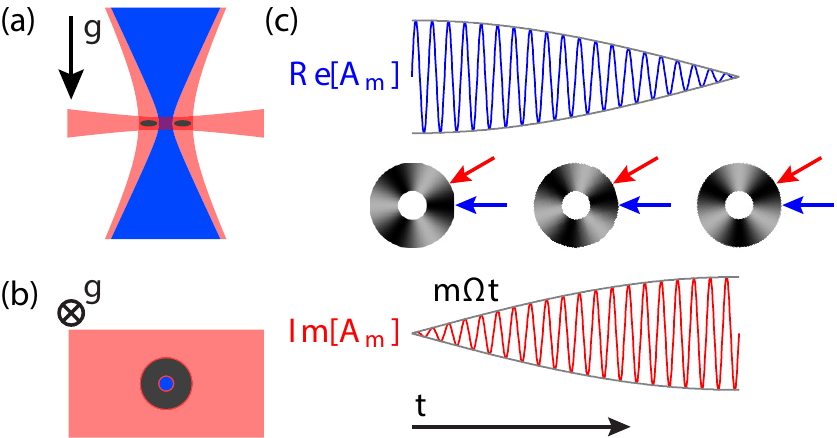}
\caption{A toroidal optical dipole trap for a $^{87}$Rb Bose-Einstein condensed gas is formed by intersecting three light fields, as indicated by the side (a) and top (b) views of the trap, with the direction of gravitational acceleration indicated. A standing acoustic wave oscillates initially at its eigenfrequency (c) with fast density oscillation at the antinode ($\mathrm{Re}[A_m]$, blue arrow and curve) and no amplitude at the node ($\mathrm{Im}[A_m]$, red arrow and curve).  Rotation induces a signal in $\mathrm{Im}[A_m]$, whose envelope (gray) increases in proportion to the rotation rate and mode number. Density shifts should not alter this envelope.\label{fig:scheme}}
\end{center}
\end{figure}

In our experiment, we prepare a degenerate, spin-polarized  gas of $^{87}$Rb atoms in a far-detuned optical dipole trap, following a procedure similar to that of Ref.\ \onlinecite{Lin2009}.  We then transfer the atoms into an overlain toroidal optical potential formed by the intersection of three light beams (Fig.~\ref{fig:scheme}(a), (b)).  An attractive light sheet, with an optical power of 12 mW, a wavelength of $\lambda = 836$ nm, and an elliptical focus with $1/e^2$ radii of $10.5\;\mu\mathrm{m}$ and $400\;\mu\mathrm{m}$, confines atoms to a horizontal plane. An annular potential, created by coaxial attractive ($400\;\mu\mathrm{W}$, $\lambda{=}830\;\mathrm{nm}$, $1/e^2$ radius of $26\;\mu\mathrm{m}$) and repulsive ($3\;\mathrm{mW}$, $\lambda{=}532\;\mathrm{nm}$, $1/e^2$ radius of $11\;\mu\mathrm{m}$) vertically propagating light beams provides in-plane confinement.  To maximize the stability of the optical system, we deliver the coaxial vertical beams via the same large-mode-area fiber.  We adjust the alignment and beam radii at the trap location using a telescope with adjustable axial and lateral chromatic shifts.  The optical potential minimum lies in a circle of radius $16\;\mu\mathrm{m}$, about which the atoms experience radial and vertical trap frequencies of $(\omega_z, \omega_r) = 2\pi\times (260, 86) \;\mathrm{Hz}$.  Evaporative cooling from the optical potential yields samples with $8\times 10^4$ condensed atoms with a $20\%$ thermal fraction, Thomas-Fermi radii of $6\;\mu\mathrm{m}$ and 1.5 $\mu$m in the radial and vertical directions, respectively, and a chemical potential of $\mu = h \times 700\;\mathrm{Hz}$. Similar optical ring traps have been demonstrated previously, albeit with different optical setups~\cite{Henderson2009,Ramanathan2011,Moulder2012}.

The excitations of a cylindrically symmetric medium may be characterized by the integer azimuthal quantum number $m$, as well as the transverse (radial and axial) quantum numbers. We excite a specific superposition of the $\pm m$ lowest-transverse-order sound modes by applying an additional optical potential to the trapped BEC that establishes an initial density modulation with high spatial overlap with the selected mode.  To create this potential, we illuminate a chrome optical mask of an $m$-fold propeller pattern (Fig.~\ref{fig:freqVsMode}(e), icons) with a $400\;\mu\mathrm{W}$, $\lambda{=}532\;\mathrm{nm}$ light beam.  The masked light is then imaged onto the plane of the trapped atoms, using a 1:10 imaging system with a resolution of $6\,\mu\mbox{m}$, imposing a repulsive optical potential while the atoms are evaporatively cooled to the final atom number and temperature.  The light is then suddenly extinguished, allowing the perturbed condensate to evolve freely.

After a variable hold time $t$, we measure the \emph{in situ} condensate density with a high signal-to-noise absorption imaging protocol. We first reduce the optical density of the gas by exciting only $10\%{-}25\%$ of the atoms to the $|F{=}2\rangle$ ground hyperfine state with a short (20 $\mu$s), weak (2\% of saturation intensity) light pulse detuned by about 5 linewidths from the $|F{=}1\rangle \rightarrow |F^\prime{=}2\rangle$ D2 transition~\cite{Ramanathan2012}. Detuning the light from resonance makes the sample optically thin and ensures that it is uniformly excited. We then apply a resonant probe pulse to the cycling transition for $50\;\mu\mathrm{s}$ at saturation intensity~\cite{Reinaudi2007} and image the unscattered light onto a CCD camera. We found this approach to give greater sensitivity than dispersive imaging.

\begin{figure}[tb!]
\begin{center}
\includegraphics[scale=1]{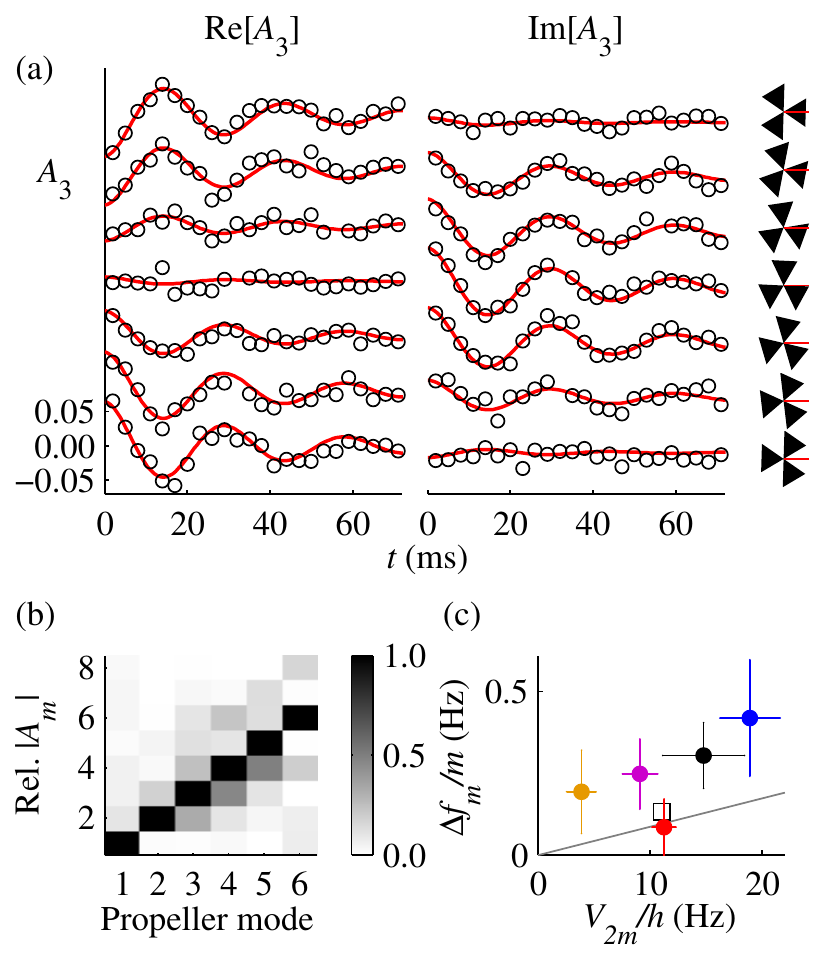}
\caption{(a) Real (left) and imaginary (right) components of $A_3$ oscillate in time. In each row, relative amplitudes correspond, with $2\pi/3$ periodicity, to the angular orientation of the excitation mask (schematic, far right). Data are offset for clarity.  (b) Relative response of $A_m$ as a function of mode number for each propeller. The applied mode always has the strongest response, but neighboring modes are weakly excited by a slight misalignment of the mask. Each column is scaled to its peak response. (c) Frequency splitting $\Delta f_m$ of mode $m$ versus the trap distortion $V_{2m} = 2 \mu |A_{2m}|$. The gray line is the expected mean-field result $\Delta f_m/m = (V_{2m}/h) (\xi/1.9r)$ for distortions in the anharmonic waveguide. The closed circle data point is from the simulated data of Section~\ref{sec:sims}. \label{fig:multiplePlots}}
\end{center}
\end{figure}

\begin{figure}[tb!]
\begin{center}
\includegraphics[scale=1]{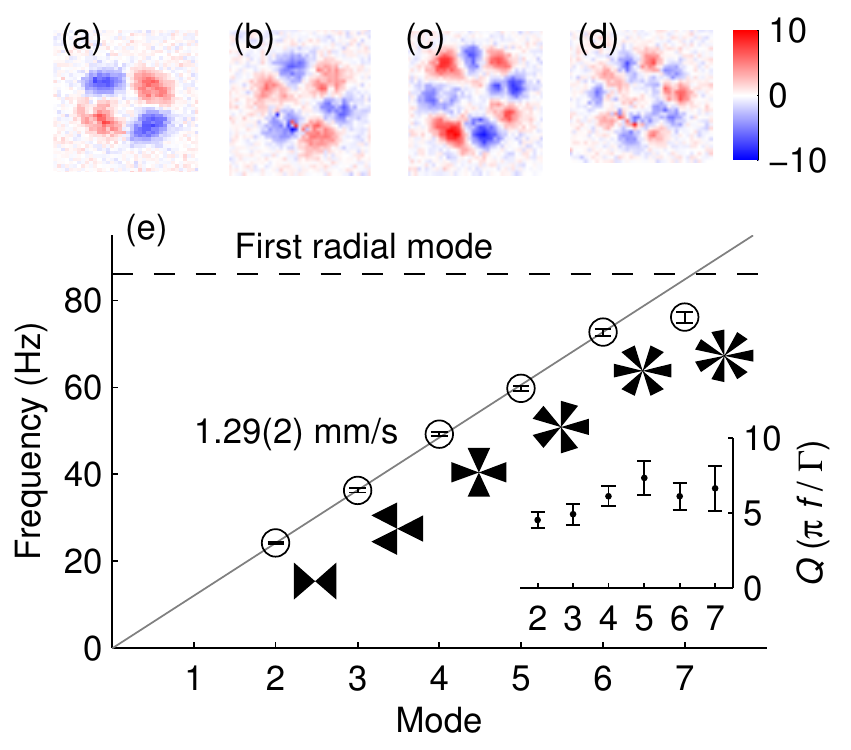}
\caption{Spatial profiles of the $m{=}2$ (a) through $5$ (d) eigenmodes are calculated from each pixel's temporal record. Each profile shows a field of view of $90\,\mu\mathrm{m}\times 90 \,\mu\mathrm{m}$. Color scale is the peak column-density modulation amplitude $\delta \tilde n(x,y)$ as defined in the text. (e)  Eigenfrequencies show the expected linear trend for modes $m{=}2{-}6$, shown as the diagonal gray line. A large negative shift of mode $m{=}7$ is expected from coupling to the first radial mode (dashed black line). The inset shows the quality factor $Q$ for each mode. \label{fig:freqVsMode}}
\end{center}
\end{figure}

The acoustic excitations reveal themselves as oscillating standing waves of the condensate density. From the measured column density distribution $\tilde n(x, y)$, we extract the azimuthal spatial Fourier coefficient
\begin{equation}
A_m = \frac{\int_{r < r_c} \dd x\,\dd y\,\tilde n\,e^{i m \phi}}{\int_{r < r_c} \dd x\,\dd y\,\tilde n},
\label{eqn:amdef}
\end{equation}
where $\phi$ is the azimuthal angle about the trap center in the imaged $x$-$y$ plane.  The integral is taken up to a cutoff radius $r_c = 40\;\mu\mathrm{m}$, where the atomic density is zero \footnote{The center of the polar coordinates is determined by fitting each image to an ideal toroidal Thomas-Fermi with a dipole angular perturbation.}. $A_m$ is a complex number whose phase indicates the angular position of the standing wave and whose amplitude indicates the strength of the density modulation. For example, Fig.~\ref{fig:multiplePlots}(a) shows the evolution of $A_3(t)$ for condensates prepared with the $m{=}3$ mask at various initial orientations, with one datum obtained per experimental cycle.

The excitation and read-out schemes are highly selective. The $m$-fold propeller produces strong excitation of the $m$-node standing wave, along with a weak residual excitation of the $(m{\pm}1)$-node standing waves due to slight misalignment of the center of the mask to the center of the trap (Fig~\ref {fig:multiplePlots}(b)). The spatial mode pattern $\delta \tilde n(x,y)$ of the standing-wave excitation can be measured by fitting the pixel-by-pixel temporal record $\tilde n(x, y, t)$ to $\tilde n_0(x, y) + \delta \tilde n(x, y)\:e^{-\Gamma t}\cos \omega (t - t_0)$, where $\Gamma$, $\omega$, and $t_0$ are fixed from a fit to $A_m$ (Fig.~\ref{fig:freqVsMode}(a-d)).

The observed standing-wave mode frequencies agree well with their predicted values~\cite{Dubessy2012} and demonstrate the expected linear phonon dispersion relation. The acoustic eigenfrequencies of a tubular medium of length $L$, an approximation for the toroid, are $f = m c_\mathrm{eff} / L$, where $c_{\rm eff}$ is the effective speed of sound in the channel. The linear scaling is seen for the $m{=}2$ through $m{=}6$ modes, with $c_{\rm eff} = 1.29(2)\;\mathrm{mm/s}$ for $L = 2\pi \times 16.0\;\mu\mathrm{m}$ in Fig.~\ref{fig:freqVsMode}(e)\footnote{The error given for the speed of sound is the $1\,\sigma$ statistical error and does not include the systematic error on the imaging system calibration.}.  The $m{=}7$ mode is perturbed by the first excited radial mode, as expected when the acoustic wavelength approaches the radial extent of the gas.  In particular, the mode frequency is reduced as the mode-pattern becomes concentrated at the outer edge of the ring, increasing the effective length and lowering $c_{\rm eff}$.

We observe a damping rate that increases with mode number, giving a quality factor that is roughly independent of mode number for modes 2-7 (inset of Fig.~\ref{fig:freqVsMode}(e)).  We note that, for our experimental conditions, free-particle (high momentum) excitations would be damped within a distance of around $d = (n \sigma)^{-1} = 14 \, \mu\mbox{m}$, shorter than that traversed by the phonon excitations studied in this work; here, $n = 10^{14} \, \mbox{cm}^{-3}$ is the peak condensate density and $\sigma = 7 \times 10^{-12} \, \mbox{cm}^2$ is the scattering cross section.

The measured damping rates are larger than those expected from Landau damping, which is a form of thermal de-excitation.  The invariance of the excitation quality factor with mode number is predicted for Landau damping in a homogeneous gas at temperatures greater than the chemical potential~\cite{Fedichev1998, Dalfovo1999}. However, our system's anisotropic potential and near-equality of the thermal energy and chemical potential ($k_B T \approx 0.9 \mu$) place it outside the regime in which Landau damping has been studied and is understood.  Indeed, the damping rate we observe is a factor of $3-6$ times higher than would be predicted by the Laudau theory at our estimated temperature of $30\;\mathrm{nK}$ \cite{smref}.  We can safely rule out four-wave mixing processes as the source of the high damping rate in our system as we have verified that the damping rate does not depend on the amplitude of the standing wave. Finally, we note that in unrelated work in the same system we have found long-lived vortices formed by rapidly cooling the gas into the BEC phase. It is possible that vortices produced inadvertently in the annulus of our ring contribute to the observed damping, and that reducing their number would enhance the lifetime of our acoustic oscillations.

\begin{figure}[tb!]
\begin{center}
\includegraphics[scale=1]{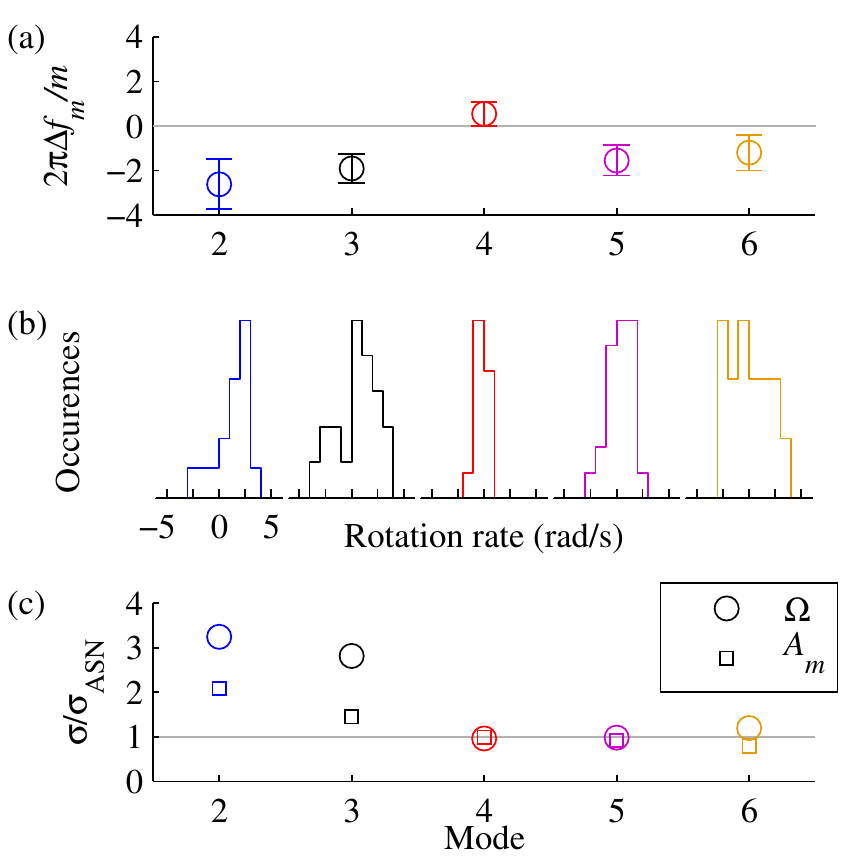}
\caption{(a) The frequency splitting is determined by fitting, on average, 500 images and 18 orientations of the optical mask per mode to a three-mode dynamical model. Error bars are determined from a jackknifing procedure that excludes a single orientation from the dataset. The units here, in rad/s per mode number, can be compared directly to the measured rotation rate. (b) Rotation estimates are binned in $1\;\textrm{rad/s}$ intervals. Each value is the result of a fit to the rotation rate for ${\sim}30$ points taken at each orientation of the optical mask. (c) The standard error $\sigma$ of the rotation estimates (closed circles) and noise in each component of $A_m$ (open squares) are close to the atom-shot-noise limit $\sigma_\mathrm{ASN}$ for modes $m{=}4{-}6$. Modes $m{=}2$ and $3$ show an excess of noise, likely from technical fluctuations in the toroidal potential.\label{fig:freqSplit}}
\end{center}
\end{figure}

\section{Proof-of-Concept Rotation Noise}\label{sec:epoc}

A rotation of the lab will create a precession in the orientation of the standing acoustic mode (see Sec.~\ref{sec:sims}). In a smooth ring, the standing wave orientation remains fixed in the inertial frame and precesses in the rotating lab frame as the clockwise ($-m$) and counterclockwise ($+m$) propagating modes appear to be split in frequency by the rotation. The equations of motion of the standing wave are identical to those a Foucault pendulum~\cite{Foucault1851}, where the two orthogonal standing waves (real and imaginary components of $A_m$) are identified with the in-plane ($x$ and $y$) components of the pendulum's position \cite{smref}. This situation differs from phonons in a rigid object, where a standing wave precesses at a rate dependent on the material shape and slower than the lab rotation rate~\cite{Bryan1890}.

In contrast, static trap distortions couple, rather than split, the $\pm m$ modes.  Owing to such distortions, the degeneracy between the sine- and cosine-like superpositions of the $\pm m$ modes is broken, and the $\pm m$ modes themselves are no longer the eigenmodes of the system. For instance, in a ring of radius $r$, a static potential perturbation with azimuthal dependence $V(\phi) = \mathrm{Re}[\sum_m V_m e^{i m \phi}]$ yields (to first order in $V_m$) a frequency splitting between the two $m$-node standing acoustic modes of
\begin{equation}\label{eqn:deltaf}
\Delta f_m \propto (|V_{2m}|/h) (\xi/\lambda),
\end{equation}
where $\xi \propto n^{-1/2}$ is the healing length~\cite{Dalfovo1999}, $n$ is the atomic density, and $\lambda$ is the acoustic wavelength.  The effects of such distortions are thus suppressed by the high density of the atomic medium \cite{smref}; for our system we observe a suppression factor of $(\Delta f_m/m)/(V_{2m}/h) \simeq 1/50$ (Fig.~\ref{fig:multiplePlots}(c)) by comparing the measured frequency splitting with trap distortions inferred from the column density $V_2m = 2 \mu |A_{2m}^\mathrm{static}|$.

While the acoustic modes are thereby affected both by rotation and by static trap distortions, the separate effects of each of these influences are distinct, akin to the separate effects of circular and linear birefringence, respectively, on the polarization of light. To quantify each effect, we excite each mode $m{=}2$ through $6$ at a large number of angles (a small selection is shown in Fig.~\ref{fig:multiplePlots}(a)), and fit the data at each $m$ to a three-mode model (unperturbed superfluid and its $\pm m$ acoustic excitations) of the temporal evolution described by the Hamiltonian
\begin{equation}
\begin{split}
H_m & =  \hbar \omega (a_x^\dagger a_x + a_y^\dagger a_y + 1)\\
&  + \hbar \pi \Delta f_m (a_x^\dagger a_x - a_y^\dagger a_y) \\
&  + i \hbar m\Omega (a_x^\dagger a_y - a_y^\dagger a_x),
\label{eqn:hamiltonian}
\end{split}
\end{equation}
where $a_x$ and $a_y$ are annihilation operators for the two non-rotating, sine- and cosine-like standing wave eigenmodes. From this fit, we obtain the eigenmode principal axes selected by the static trap distortion (Fig.~\ref{fig:freqSplit}, top) and the eigenmode frequencies $\omega_{x/y} = \omega \pm \pi \Delta f_m$, as well as the rotation rate $\Omega$. The simplified three-mode description of our system is consistent with the dynamics predicted by numerically integrating the Gross-Pitaevskii equation for our system parameters in the limit of low amplitude excitations of a single $m$-mode, as demonstrated in Section \ref{sec:sims}.

To extract a rotation rate, we fit the data at each mask orientation (${\sim}30$ points, corresponding to a row in Fig.~\ref{fig:multiplePlots}(a)) to a prediction of the sound wave evolution in a rotating frame. The extracted rates are plotted as a histogram in Fig.~\ref{fig:freqSplit}(b). The only free parameters are the rotation rate and the amplitude of the initial excitation. Mechanical properties of the system such as the frequency, frequency splitting, and phase are fixed from fits to the rest of the data set.

\begin{figure*}[tb!]
\begin{center}
\includegraphics[scale=1]{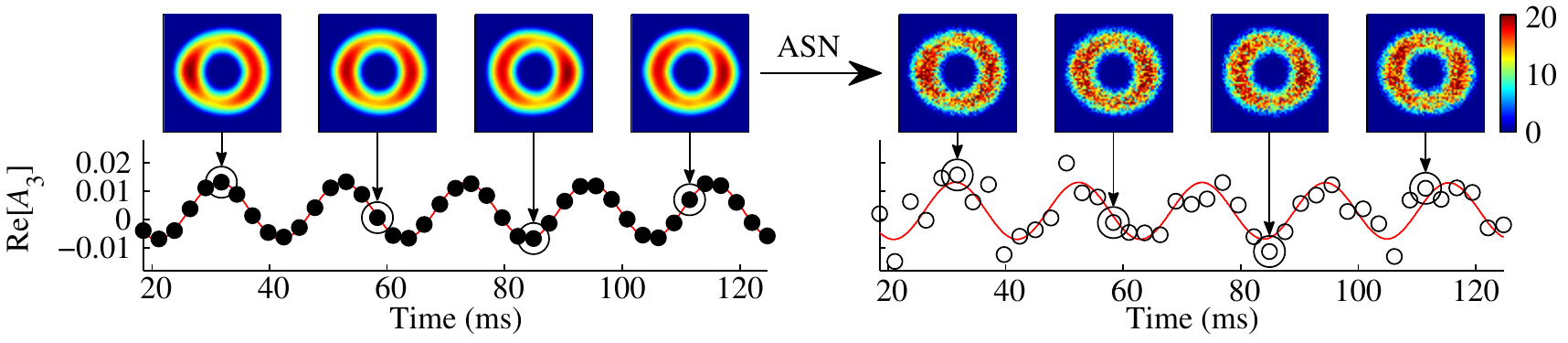}
\caption{$A_3$ is extracted from snapshots of the simulated condensate column density both with (right) and without (left) the addition of simulated atom shot noise. For each trap configuration, 300 samples over 793 ms at each of 5 angles of the excitation propeller are analyzed using the three-mode model. Oscillation of the $m{=}3$ mode is exaggerated by a factor of 4 in the included images, for clarity. \label{fig:simscheme}}
\end{center}
\end{figure*}

The measured rotation noise is at the level of 1 rad/s.  Considering that these measurements are obtained in around 30 repetitions of the experiment, each of which has an average measurement time of tens of milliseconds, this rotation noise can also be expressed as approximately 1 $\mbox{rad} \, \mbox{s}^{-1} / \sqrt{Hz}$ where we do not account for the low duty cycle of the experiment.  This measurement noise matches well with the expected atom-shot-noise limit (Fig.~\ref{fig:freqSplit}, bottom). In imaging the distribution of $N$ uncorrelated atoms, atom-shot-noise yields an uncertainty $\Delta A_m = (2 N)^{-1/2}$ in extracting either the real or imaginary component of $A_m$ in a single run.  The atom shot noise limits the rotation signal to an uncertainty of
\begin{equation}\label{eq:ASN}
 \Delta \Omega_\mathrm{ASN} = \frac{\alpha \Gamma}{m A \sqrt{N N_r}},
 \end{equation} where $\Gamma$ is the decay rate of the standing wave, $A$ is the fractional density modulation excited in operating the interferometer (measured by the Fourier amplitude $A_m$ defined in Eq.\ \ref{eqn:amdef}), and $N_r$ is the number of independent experimental realizations. The prefactor $\alpha$ depends on how measurements are distributed in time; here, $\alpha=3.2$ for evolution times sampled uniformly between $0$ and $2/\Gamma$ \cite{smref}.  For measurements using the higher-order sound modes, the reduced technical noise in our images at higher spatial frequencies allows us to achieve the atomic shot-noise limit (Fig.\ \ref{fig:freqSplit}(c)).  Unfortunately, because the quality factor of acoustic standing-wave oscillations is found to be roughly independent of the mode number, the use of higher order modes does not improve sensitivity in our apparatus beyond overcoming technical noise.

The fundamental sensitivity due to atom shot noise in our device, based on interfering phonons, takes a very similar form to a conventional free-space atom interferometer, based on interfering atoms. In the latter case, the noise can be written as
\begin{equation}
\Delta \Omega_\mathrm{free} = \frac{T^{-1}}{4\pi (L/\lambda) \sqrt{N}},
\end{equation} where $L$ is the distance between interaction regions, $\lambda$ is the wavelength of the optical or material grating used as an atomic beam splitter, and $T$ is the atomic travel time between beam splitters~\cite{Gustavson2000}.  Written this way, our acoustic interferometer acts as a device that measures how far a feature of azimuthal size $\lambda/L \sim 1/m$ rotates over a time $T \sim 1/\Gamma$.

\section{Mean-field Numerics}\label{sec:sims}

The short lifetime of acoustic modes in our apparatus makes evaluating the limits and efficacy of the three-mode model of Eq.\ \ref{eqn:hamiltonian}, and our sensing scheme in general, difficult. Thus, we explore the feasibility of our scheme in the absence of anomalous damping by simulating it numerically. These results validate the essential points of our proposal, namely that interferometers based on collective excitations within gaseous superfluids suppress the impacts of common sources of systematic bias and, being irrotational, provide an absolute non-rotating frame of reference.

We modeled our system using the Gross-Pitaevskii (GP) equation, which we discretized via a symmetric split-step Fourier method~\cite{Feit1982,Taha1984} on a $N_x {\times} N_y {\times} N_z {=} 64{\times} 64{\times} 8$ grid and with an adaptive temporal step size, giving spatial accuracy to infinite order for length scales above ${\sim}0.6\;\mathrm{\mu m}$ and local temporal accuracy to $\mathcal{O}(\Delta t^4)$, with $\Delta t$ the contemporaneous step size. Error thresholds were chosen such that accumulated errors throughout a typical simulated experiment were negligible.

Each simulated experiment began by finding the ground state of $10^5$ $^{87}\mathrm{Rb}$ atoms in a non-rotating potential, including an $m$-fold repulsive propeller and a simulated 3-beam optical dipole trap, possibly with static inhomogeneity, via evolution in imaginary time. From this starting point, we excited acoustic oscillations by suddenly removing the $m$-fold propeller and propagating the system in real time, rotating the simulated lab at a rate $\Omega_0$. We implemented the rotation of the lab by rotating the trapping potentials, including any static inhomogeneity, with respect to the non-rotating frame of the simulation.

We chose a non-rotating ground state as our starting point because of its compatibility with the split-step Fourier method. For small rotation rates, the difference between rotating and non-rotating ground states is small.  This difference manifests itself in our simulations as an additional acoustic oscillation that we would not expect to be present in our experimental apparatus. These acoustic oscillations can be observed on their own by running simulated experiments in which the propeller pattern is not turned off immediately. Owing to its irrotational nature, the superfluid is not dragged by the rotating propeller pattern.

\begin{figure}[tb!]
\begin{center}
\includegraphics[scale=1]{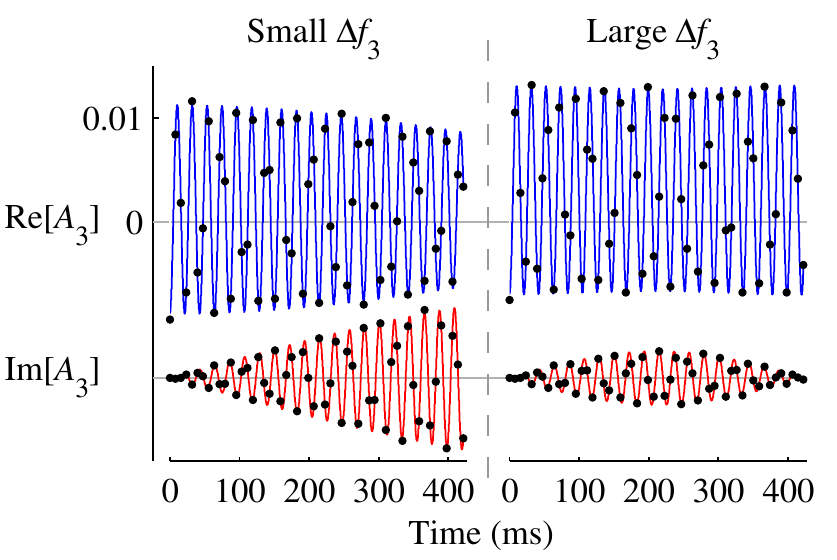}
\caption{Sample simulated data (no atom shot noise) with an initial excitation along an $m{=}3$ principal axis at two different values of frequency splitting $\Delta f_3$. The rotation rate in both datasets is $\Omega_0 {=} 0.1\; \mathrm{Hz}$ and the confining traps are the same except for differing 5-fold ($V_5$) and 6-fold ($V_6$) perturbations of $h\times11\; \mathrm{Hz}$ (small $\Delta f_3$) and $h\times 59\; \mathrm{Hz}$ (large $\Delta f_3$). \label{fig:simdata}}
\end{center}
\end{figure}

To produce data analogous to the images from our experimental apparatus, we saved snapshots of the simulated condensate column density, integrated along the axial dimension, at various times. To understand and verify our assumptions about the role of atom shot noise, we derived noisy measurements of the condensate column density from these snapshots by adding Gaussian noise at a level consistent with the atom shot noise limit for images including $2{\times1}0^4$ atoms. We extracted $A_m$ from both the noisy and noiseless snapshots in the rotating lab frame (Fig.\ \ref{fig:simscheme}), mirroring the experimental procedure and analysis presented in Section\ \ref{sec:epoc}.

In our simulations, we chose to focus on excitations of an $m{=}3$ propeller in a trapping potential with parameters similar to those of our experimental apparatus, but with additional variable 5-fold and 6-fold angular perturbations of the trapping potential. The simulated light sheet, modeled numerically as if formed by a focused Gaussian light beam, naturally includes a moderate 2-fold inhomogeneity.

The sample data shown in Fig.\ \ref{fig:simdata} demonstrate important qualitative features of the sensing scheme and the three-mode model (Eq.\ \ref{eqn:hamiltonian}). For small $\Delta f_3$ (small 6-fold perturbation in trapping potential, $V_6$), the rotation $\Omega_0{=}0.1\;\mathrm{Hz}$ causes an excitation that is initially along one principal axis to rotate into the orthogonal axis at a rate of $3 \Omega_0$. When $\Delta f_3$ is large (large $V_6$), the excitation remains pinned along its initial axis. The offset in $\mathrm{Re}[A_3]$ in the data for large $\Delta f_3$ is due to the larger 5-fold trap perturbation, $V_5$, which gives a small static 3-fold perturbation when combined with the 2-fold perturbation of the simulated light sheet (one of the three optical trapping beams, as described in Sec.\ \ref{sec:setup}).

\begin{figure}[tb!]
\begin{center}
\includegraphics[scale=1]{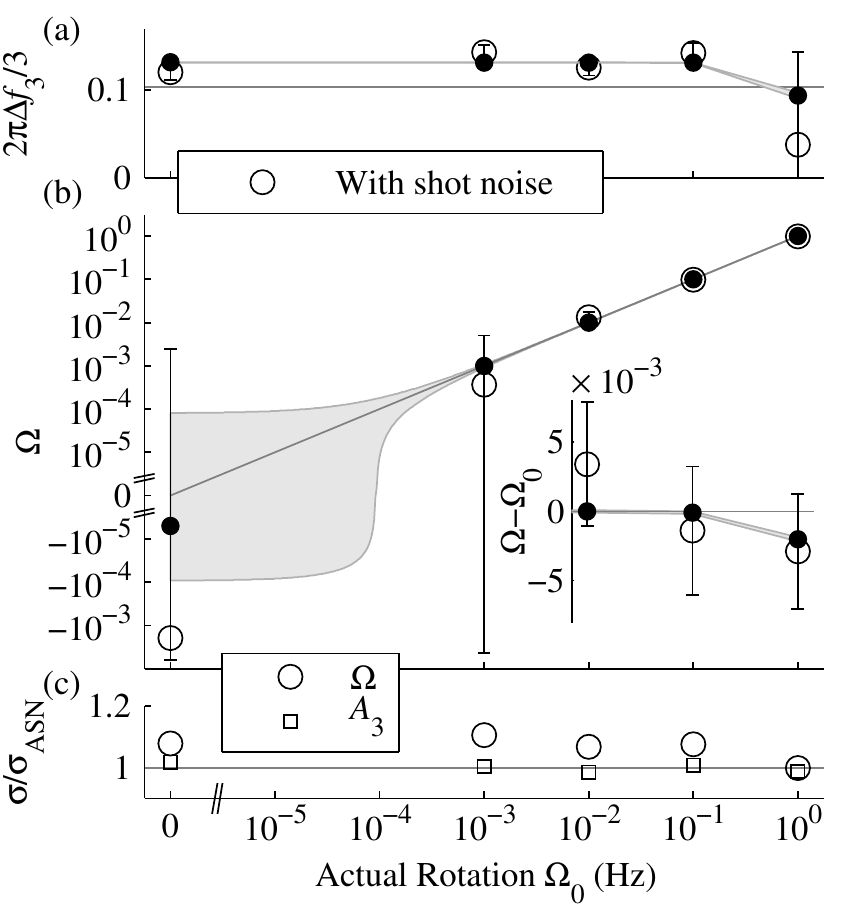}
\caption{ (a) Frequency splitting, (b) rotation rate $\Omega$, and (inset) $\Omega{-}\Omega_0$ residuals of the $m{=}3$ mode extracted from simulated data with (open circles) and without (closed circles) atom shot noise. Error bars apply to the open circle data points. The filled gray area delineates the uncertainty in parameters extracted from the noiseless data and represents the limits of the three-mode description of the full mean-field dynamics in the simulated trap and with the simulated excitation, both of which were set to match real experimental parameters. The three-mode model distinguishes between rotation and frequency splitting, and can extract both from the same data stream.  The measured $2\pi \Delta f_3/3{=}0.13$ is close to the value 0.10  expected based on the static trap perturbation $V_6$ (Eq.\ \ref{eqn:deltaf}). (c) The standard deviation of $\Omega$ and $A_m$ extracted from noisy snapshots  is consistent with the expected fit-free atom shot noise limit (Eq.\ \ref{eq:ASN}).  \label{fig:simresult}}
\end{center}
\end{figure}

Our analysis scheme based on the three-mode model, applied to the results of our full numerical simulation, is able to extract the mechanical parameters of the oscillator, including the mode frequency splitting, as well as the rotation rate. Figure \ref{fig:simresult} shows the results of fitting the three-mode model to  300 snapshots over 793 ms at each of 5 angles of a 3-fold propeller. All model parameters were fit simultaneously, with error bars determined by a delete-one jackknife procedure~\cite{MILLER1974}.  Errors obtained via this procedure were consistent with errors obtained directly from fits.

The extracted rotation rates are well-calibrated for a large range of rotation values. As expected from the irrotational flow of the superfluid, small rotation rates can be measured absolutely, without bias drift or pinning. For large rotation rates, the extracted values are consistent with the actual rotation rate beyond 0.1 Hz. At 1 Hz, the measured rotation rate slips to 0.998 Hz. At this rotation rate, the true ground-state of the rotating system has a high probability of entraining a vortex, but our simulations begin with a condensate in the non-rotating ground state; the additional acoustic oscillations caused by the large deviation of our simulated system from the true ground-state at this high rate of rotation are, evidently, no longer negligible and they likely account for our failure to accurately measure rotations in this limit. While physical systems based on this implementation would not suffer the same limitation, they would likely need to account for the probability of entraining a quantized vortex in the model used to extract rotation to be accurate at high rotation rates.

Uncertainty in the values of parameters of the three-mode model persist, even in the absence of noise, as indicated by the filled gray areas in Fig.\ \ref{fig:simresult}. This residual uncertainty indicates the degree to which the three-mode model accurately describes the temporal evolution of the acoustic excitation in the mean-field limit. Our simulations have shown that the three-mode model works best for low-amplitude excitations wherein the excitation propeller excites only the lowest order of the desired $m$-fold perturbation.

\section{Conclusion}
For a trapped atom sample to reach sensitivities competitive to free-space interferometers, short wavelengths (higher $m$), large atom numbers, long propagation times, and very selective excitation are required. In our current setup, both atom number and $m$ are limited by the small size of our ring. However, rings have been demonstrated with $\approx 10^4$ greater enclosed area~\cite{Gupta2005}. Further, collective modes of BECs have been observed with temperature-limited quality factors ten times greater than those reported here~\cite{Jin1997,Stamper-Kurn1998}.  Fundamental zero-temperature damping of these modes, Beliaev damping, should allow for subhertz damping rates for wavelengths a few times greater than the healing length~\cite{Katz2002}.

More importantly, we emphasize that collective excitation interferometry has applications beyond rotation sensing with phonons. Other collective excitations and geometries could be employed in a diverse range of sensors. For example, magnons in a ferromagnetic spinor BEC are predicted to show free-particle behavior and could form the basis of a compact, short-range interferometric magnetic sensor.

We thank A.\ \"Ottl and M.\ Solarz for help with designing and building the cold atom setup and G.\ Dunn and S.\ Lourette for technical assistance. This work was supported by the Defense Advanced Research Project Agency (Grant No. 49467-PHDRP) and the Defense Threat Reduction Agency (Contract No. HDTRA1-09-1-0020). GEM acknowledges support from the Hertz Foundation.

\bibliography{sound,smref-arxiv}


\clearpage
\newpage



\section*{Supplementary material}

# Collective Excitation Interferometry with a Toroidal Bose-Einstein Condensate: Supplemental Material

G. Edward Marti,[1, *] Ryan Olf,[1, †] and Dan M. Stamper-Kurn[1, 2]

[1]*Department of Physics, University of California, Berkeley, California 94720, USA*
[2]*Materials Sciences Division, Lawrence Berkeley National Laboratory, Berkeley, California 94720, USA*

## S1. EQUATION OF MOTION OF $A_m$

The motion of the sound waves with mode number $m$ is identical to that of an anisotropic two-dimensional oscillator (of mass $M$), with natural frequencies of $\omega \pm \delta$, that is rotating at the rate $m\Omega$, where $\Omega$ is the lab-frame rotation rate for the sound-wave system. The classical Hamiltonian for this oscillator system is

$$H = \frac{p_x^2}{2M} + \frac{p_y^2}{2M} - m\Omega(xp_y - yp_x)$$
$$+ \frac{M}{2}(\omega+\delta)^2 x^2 + \frac{M}{2}(\omega-\delta)^2 y^2.$$

We can define (unperturbed) annihilation operators $a_x$ and $a_y$ for the two non-rotating eigenmodes to find the quantum Hamiltonian of Eq. (3) in the main text. The classical result can be recovered by considering a three state system, $|0,0\rangle$, $|0,1\rangle$ and $|1,0\rangle$, written in the $|n_x, n_y\rangle$ basis. The Hamiltonian can then be represented by a matrix,

$$H \Leftrightarrow \hbar \begin{pmatrix} \omega+\delta & -i\Omega & 0 \\ i\Omega & \omega-\delta & 0 \\ 0 & 0 & 0 \end{pmatrix},$$

and the two position operators are,

$$x \propto \begin{pmatrix} 0 & 0 & 1 \\ 0 & 0 & 0 \\ 1 & 0 & 0 \end{pmatrix} \qquad y \propto \begin{pmatrix} 0 & 0 & 0 \\ 0 & 0 & 1 \\ 0 & 1 & 0 \end{pmatrix}.$$

It is sufficient to solve for the response of the system when kicked along an eigenaxis. Linearity will let us combine solutions to account for an arbitrary kick. We define $A_{\sigma\nu}$ as the response along the $\nu$ axis to a system initially oscillating along the $\sigma$ axis. Hence, if the pendulum starts along $x$, then we find:

$$A_{xx} = \cos(\omega t)\cos(\Omega' t) - \frac{\delta}{\Omega'}\sin(\omega t)\sin(\Omega' t)$$
$$A_{xy} = \frac{\Omega}{\Omega'}\cos(\omega t)\sin(\Omega' t), \qquad (S1)$$

---
* emarti@berkeley.edu
† ryanolf@berkeley.edu

where $\Omega' = \sqrt{\Omega^2 + \delta^2}$. For an oscillator kicked along the $y$ direction,

$$A_{yx} = -\frac{\Omega}{\Omega'}\cos(\omega t)\sin(\Omega' t)$$
$$A_{yy} = \cos(\omega t)\cos(\Omega' t) + \frac{\delta}{\Omega'}\sin(\omega t)\sin(\Omega' t).$$

We emphasize that these results, as well as the equations of motion calculated for Bogoliubov modes, are identical to the classical ones.

## S2. DERIVATION OF MEAN-FIELD SPLITTING

Consider a one-dimensional circular waveguide. For uniform radial perturbations, we estimate the potential as a function of radius $r$ and azimuthal angle $\phi$ as $V(r,\phi) = V(r) + \mathrm{Re}[\sum_m V_m e^{im\phi}]$. The effective (channel) speed of sound is

$$c_{\mathrm{eff}} = \sqrt{\frac{\mu - V(\phi)}{\zeta\, m_{\mathrm{Rb}}}} \approx c_0\left(1 - \frac{V(\phi)}{2\mu}\right), \qquad (S2)$$

where $\mu$ is the chemical potential, $m_{\mathrm{Rb}}$ is the atomic mass, and $\zeta$ is a constant that depends on the form of $V(r)$ (1 for a square potential, 2 for a harmonic potential, and $\sim 1.8$ for our anharmonic potential). By applying perturbation theory to the sound equation, we find that the two $m$ modes have the perturbed frequencies

$$\omega_m = \omega_{m,0}\left(1 \pm \frac{|V_{2m}|}{4\mu}\right), \qquad (S3)$$

with the unperturbed frequency $\omega_{m,0} = m c_{\mathrm{eff}}/r$. We report the frequency $f_m = \omega_m/2\pi$ and the frequency splitting $\Delta f_m = \delta/\pi$. In Fig. 3(c) of the main text, the theory line can be rewritten as

$$\frac{\Delta f_m}{m} = V_{2m}\frac{f_m/m}{2\mu} = \frac{V_{2m}}{h}\frac{\xi}{r\sqrt{2\zeta}}$$

where $\xi = \hbar/\sqrt{2\, m_{\mathrm{Rb}}\mu}$ is the healing length [1].

## S3. DERIVATION OF ATOM-SHOT-NOISE LIMIT

We first derive the atom-shot-noise limit in measuring $A_m$, and then in the rotation rate, $\Omega$. The two quadratures of $A_m$ have equal noise, so without loss of generality,

we calculate only the noise in the real quadrature. For each pixel $i$ we detect $n_i$ atoms with an atom-shot-noise variance of $\bar{n}_i$. For a large number of atoms ($N \gg n_i$),

$$\text{Re}[A_m] = \frac{\sum_i n_i \cos m\phi_i}{\sum_i n_i}$$

$$\text{var Re}[A_m] \approx \frac{1}{N^2} \sum_i \text{var}(n_i \cos m\phi_i)$$

$$= \frac{1}{N^2} \sum_i \bar{n}_i \cos^2 m\phi_i$$

$$\approx \frac{1}{2N}.$$

To estimate $\Delta\Omega_{\text{ASN}}$, we make the short-time, small rotation rate approximation, $\Omega t \ll 1$ and $\delta t \ll 1$, for which a kick along the real axis leads to a response along the imaginary axis. The relevant signal is the imaginary component of $A_m$,

$$\text{Im}[A_m] = A\, e^{-\Gamma t} \cos \omega t \sin \Omega m t.$$

In each iteration of the experiment, the atom-shot-noise uncertainty in $\Omega$ is

$$\Delta\Omega(t) = \frac{\text{std Im}[A_m]}{|\partial\text{Im}[A_m]/\partial\Omega|} = \frac{1/\sqrt{2N}}{Amt\, e^{-\Gamma t}|\cos\omega t|}.$$

We choose to uniformly sample our data on an interval $t \in [0, b/\Gamma]$, where $b$ is between 2 and 3. For a large number of samples, the linear least-squares estimate of the error is

$$\Delta\Omega_{\text{ASN}} = \left[\sum_i \Delta\Omega(t_i)^{-2}\right]^{-1/2}$$

$$= \left[\frac{\Gamma}{b} \int_0^{b/\Gamma} dt\, 2NA^2 m^2 t^2 e^{-2\Gamma t} \cos^2 \omega t\right]^{-1/2}$$

$$= \frac{\Gamma}{Am\sqrt{N}} \sqrt{\frac{8b}{\int_0^{2b} u^2 e^{-u}\, du}}$$

$$\approx (3.2 \text{ to } 3.6) \frac{\Gamma}{Am\sqrt{N}}.$$

## S4. THERMAL AND ZERO-TEMPERATURE SCATTERING

Landau damping, collisions with thermally excited collective modes, is expected to be the main form of damping of sound in our system. A collective excitation in a homogeneous Bose-Einstein condensate at chemical potential $\mu$ and temperature $T$ has a quality factor $Q$ independent of mode number and inversely proportional to temperature,

$$Q = \frac{\omega}{2\Gamma} = \frac{4}{3\pi} \frac{\hbar c}{k_B T a},$$

where $\omega$ is the (real) mode frequency, $\Gamma$ is the damping rate (in radians per second), $c$ is the speed of sound and $a$ is the s-wave scattering length. This approximate assumes $\hbar\omega \ll \mu \ll k_B T$, which does not hold in our system. Regardless, we find this description qualitatively agrees with our data but predicts a quality factor 3−6 times larger than observed. Other regimes, such as when $\mu \approx k_B T$, have been considered theoretically [2], and the predicted damping rates under such conditions area similarly smaller than what we observe experimentally.

At zero temperature, scattering of collective modes should be dominated by Beliaev scattering, a four-wave mixing process (see Ref. 3 for calculations relevant to Bose-Einstein condensates). This form of damping is characterized by a very strong dependence of the damping rate on the wavenumber, $\Gamma \sim k^5$. For wavelengths as short as one micron, much shorter than used in this work, the damping rate should be below 1 s$^{-1}$ and far below what is experimentally observed.


\clearpage
\end{document}